\documentclass[journal,twocolumn]{IEEEtran}%
\usepackage{graphicx}%
\usepackage{amsmath}%
\usepackage{amsfonts}%
\usepackage{amssymb}
\begin{document}

\title{Observation of Bifurcations and Hysteresis in Nonlinear NbN Superconducting
Microwave Resonators}
\author{Baleegh Abdo, \textit{Student Member},\textit{\ IEEE}, Eran Segev, Oleg
Shtempluck, and~Eyal Buks~\thanks{This work has been submitted to the IEEE for
possible publication. This work was supported by the German Israel Foundation
under grant 1-2038.1114.07, the Israel Science Foundation under grant 1380021,
the Deborah Foundation and Poznanski Foundation.}\thanks{The authers are with
the Department of Electrical Engineering and Microelectronics Research Center,
Technion, Haifa 32000, Israel (e-mail: baleegh@tx.technion.ac.il).}}
\maketitle

\begin{abstract}
In this paper we report some extraordinary nonlinear dynamics measured in the
resonance curve of NbN superconducting stripline microwave resonators. Among
the nonlinearities observed: aburpt bifurcations in the resonance response at
relatively low input powers, asymmetric resonances, multiple jumps within the
resonance band, resonance frequency drift, frequency hysteresis, hysteresis
loops changing direction and critical coupling phenomenon. Weak links in the
NbN grain structure are hypothesized as the source of the nonlinearities.

\end{abstract}

\begin{keywords}
Bifurcations, nonlinear effects, hysteresis, NbN, microwave resonators.
\end{keywords}

%Note that keywords are not normally used for peerreview papers.

%For peer review papers, you can put extra information on the cover
%page as needed:
%\begin{center} \bfseries EDICS Category: 3-BBND \end{center}
%For peerreview papers, inserts a page break and creates the second title.
%Will be ignored for other modes.

\section{Introduction}

%The very first letter is a 2 line initial drop letter followed
%by the rest of the first word in caps.
%form to use if the first word consists of a single letter:
%\PARstart{A}{demo} file is ....
%form to use if you need the single drop letter followed by
%normal text (unknown if ever used by IEEE):
%\PARstart{A}{}demo file is ....
%Some journals put the first two words in caps:
%\PARstart{T}{his demo} file is ....
%Here we have the typical use of a "T" for an initial drop letter
%and "HIS" in caps to complete the first word.
\PARstart{N}{onlinear} effects in superconductors in the microwave regime have
been the subject of a large number of intensive studies in recent years. Most
of the attention is focused on studying one or more of the following issues:
investigating the origins of nonlinear effects in superconductors \cite{rf
residual losses Halbritter},\cite{Jerusalem}, introducing theoretical models
that explain nonlinear behavior \cite{yip},\cite{influence}, identifying the
dominant factors that manifest these effects \cite{nonlinear understanding}%
,\cite{The truth}, find ways to control and minimize nonlinear effects
\cite{dahm},\cite{oxygen content} such as, harmonic generation and
intermodulation distortions, which degrade the performance of promising
superconducting microwave applications mainly in the telecommunication area
\cite{optimization}.

Among the nonlinear effects reported in the literature associated with
resonance curves, one can find the commonly known Duffing nonlinearity which
is characterized by skewed resonance curves above certain power level,
\ appearance of infinite slope in the resonance lineshape, pronounced shift of
the resonance frequency and hysteretic behavior \cite{nonlinear
electrodynamics of sc NbN and Nb},\cite{Thermally-induced nonlinearities
surface impedance sc YBCO},\cite{suspended HTS mw resonator}. To account for
this effect, associated with the rise of kinetic inductance of
superconductors, both thermal \cite{Thermally induced nonlinear behaviour},
and weak link \cite{nonlinear electrodynamics of sc NbN and Nb} explanations
have been successfully applied. Other nonlinear effects were reported by
Portis \textit{et al.} \cite{HTS patch antenna}, where they observed notches
that develop on both sides of the frequency response of their HTS microstrip
patch antenna, accompanied with hysteresis and frequency shift, as they have
driven their antenna into the nonlinear regime. Similar results were reported
also by Hedges \textit{et al.} \cite{power dependent effects observed for sc
stripline resonator}, in their YBCO stripline resonator,\ and by \cite{mw
power handling weak links thermal effects} in their YBCO thin film dielectric
cavity. All three studies attributed the observed nonlinear behavior to abrupt
changes in the resistive loss of weak links, thermal quenching and weak link
switching to normal state.

In this study, being interested in the behavior of nonlinear resonances, we
have fabricated different NbN superconducting microwave resonators exhibiting
some unusual nonlinear effects, which to the best of our knowledge, have not
been reported before in the literature. We study the dependence of these
resonators on the injected input power level, and examine the resonance curve
behavior under different scan directions showing interesting features. To
account for our results, we consider briefly some possible physical mechanisms
that may be responsible for the observed effects.
%TCIMACRO{\TeXButton{TeX field}{\begin{table*}
%\renewcommand{\arraystretch}{1.5}
%\caption{Sputtering Parameters}
%\centering\begin{tabular} {c | c | c | c}
%\hline\bfseries Process parameter & \bfseries B1 & \bfseries B2 & \bfseries
%B3 \\
%\hline\hline\bfseries
%Partial flow ratios (Ar,N2) & (87.5\%,12.5\%)  & (75\%,25\%)  & (70\%,30\%) \\
%\hline\bfseries Base temperature  & $11\,^{\circ}\mathrm{C}$  & $11\,^{\circ
%}\mathrm{C}$  & $13\,^{\circ}\mathrm{C}$ \\
%\hline\bfseries Total pressure & $ 6.9 \cdot10^{-3} $  torr & $ 8.1 \cdot
%10^{-3} $ torr & $ 5.7 \cdot10^{-3} $ torr \\
%\hline\bfseries Discharge current & $0.36 \mathrm{A} $  & $0.55 \mathrm{A}
%$ & $0.36 \mathrm{A} $ \\
%\hline\bfseries Discharge voltage & $351 \mathrm{V} $  & $348 \mathrm{V}
%$ & $348 \mathrm{V} $ \\
%\hline\bfseries Discharge power & $121 \mathrm{W}$  & $185 \mathrm{W}
%$ & $133 \mathrm{W} $ \\
%\hline\bfseries Deposition rate & $6 \mathrm{\frac{\AA}{\sec}}
%$  & $7.8 \mathrm{\frac{\AA}{\sec}}  $ & $3.8 \mathrm{\frac{\AA}{\sec}}
%$ \\
%\hline\bfseries Thickness (t) & $2200 \AA$  & $3000 \AA$  & $2000 \AA$   \\
%\hline\bfseries Base pressure & $ 3.1 \cdot10^{-8} $  torr & $ 7.3 \cdot
%10^{-8} $ torr & $ 8 \cdot10^{-8} $ torr \\
%\hline\bfseries Target-substrate distance & $80 \mathrm{mm} $  & $90 \mathrm
%{mm} $ & $90\mathrm{mm} $ \\
%\hline\end{tabular}
%\end{table*}}}%
%BeginExpansion
\begin{table*}
\renewcommand{\arraystretch}{1.5}
\caption{Sputtering Parameters}
\centering\begin{tabular} {c | c | c | c}
\hline\bfseries Process parameter & \bfseries B1 & \bfseries B2 & \bfseries
B3 \\
\hline\hline\bfseries
Partial flow ratios (Ar,N2) & (87.5\%,12.5\%)  & (75\%,25\%)  & (70\%,30\%) \\
\hline\bfseries Base temperature  & $11\,^{\circ}\mathrm{C}$  & $11\,^{\circ
}\mathrm{C}$  & $13\,^{\circ}\mathrm{C}$ \\
\hline\bfseries Total pressure & $ 6.9 \cdot10^{-3} $  torr & $ 8.1 \cdot
10^{-3} $ torr & $ 5.7 \cdot10^{-3} $ torr \\
\hline\bfseries Discharge current & $0.36 \mathrm{A} $  & $0.55 \mathrm{A}
$ & $0.36 \mathrm{A} $ \\
\hline\bfseries Discharge voltage & $351 \mathrm{V} $  & $348 \mathrm{V}
$ & $348 \mathrm{V} $ \\
\hline\bfseries Discharge power & $121 \mathrm{W}$  & $185 \mathrm{W}
$ & $133 \mathrm{W} $ \\
\hline\bfseries Deposition rate & $6 \mathrm{\frac{\AA}{\sec}}
$  & $7.8 \mathrm{\frac{\AA}{\sec}}  $ & $3.8 \mathrm{\frac{\AA}{\sec}}
$ \\
\hline\bfseries Thickness (t) & $2200 \AA$  & $3000 \AA$  & $2000 \AA$   \\
\hline\bfseries Base pressure & $ 3.1 \cdot10^{-8} $  torr & $ 7.3 \cdot
10^{-8} $ torr & $ 8 \cdot10^{-8} $ torr \\
\hline\bfseries Target-substrate distance & $80 \mathrm{mm} $  & $90 \mathrm
{mm} $ & $90\mathrm{mm} $ \\
\hline\end{tabular}
\end{table*}%
%EndExpansion

\section{Resonators Design}

\subsection{Resonator Geometries}

\begin{figure}
[ptb]
\begin{center}
\includegraphics[
height=1.235in,
width=3.6754in
]%
{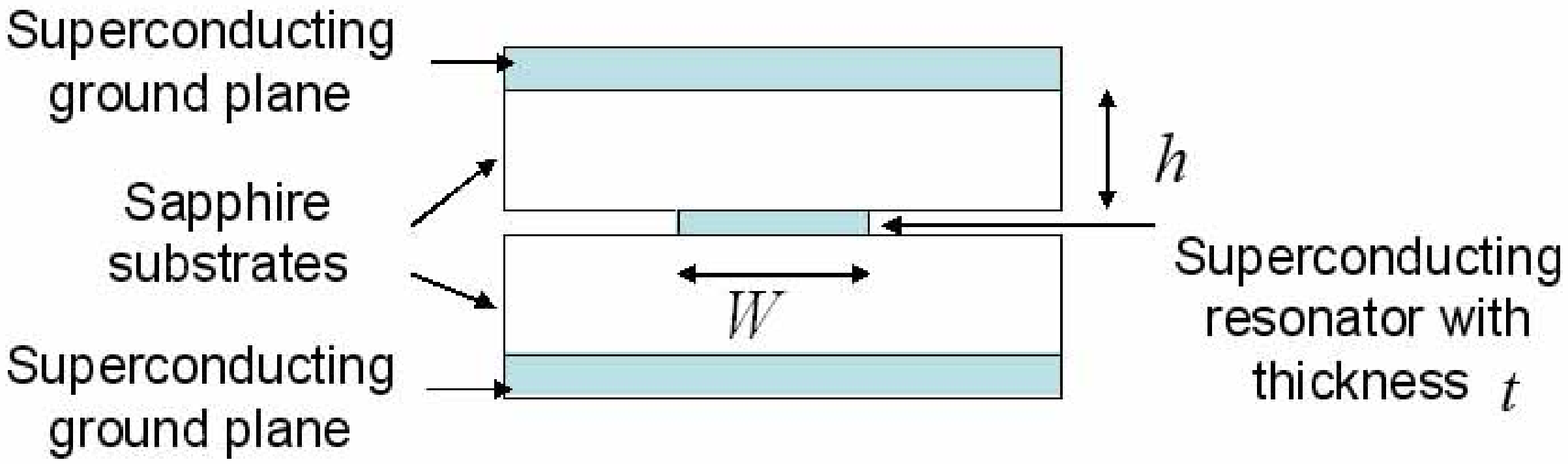}%
\caption{Stripline geometry.}%
\label{stripline_geometry}%
\end{center}
\end{figure}

The resonators were designed in the standard stripline geometry, which
consists of five layers as shown in the cross section illustration depicted in
Fig. \ref{stripline_geometry}.%

%TCIMACRO{\FRAME{ftbpFU}{3.6754in}{1.235in}{0pt}{\Qcb{Stripline geometry.}%
%}{\Qlb{stripline_geometry}}{stripline.eps}%
%{\special{ language "Scientific Word";  type "GRAPHIC";  display "USEDEF";
%valid_file "F";  width 3.6754in;  height 1.235in;  depth 0pt;
%original-width 10.0854in;  original-height 2.9525in;  cropleft "0";
%filename '../stripline.eps';file-properties "XNPEU";}}}%
%BeginExpansion
%EndExpansion

The superconducting resonator was dc-magnetron sputtered on one of the
sapphire substrates, whereas the superconducting ground planes were sputtered
on the inner covers of a gold plated Oxygen Free High Conductivity (OFHC)
Cupper Faraday package that was employed to house the resonators. The
dimensions of the sapphire substrates were $34%
%TCIMACRO{\unit{mm}}%
%BeginExpansion
\mathrm{mm}%
%EndExpansion
$ $X$ $30%
%TCIMACRO{\unit{mm}}%
%BeginExpansion
\mathrm{mm}%
%EndExpansion
$ $X $ $1%
%TCIMACRO{\unit{mm}}%
%BeginExpansion
\mathrm{mm}%
%EndExpansion
.$ The resonator geometries implemented, which we will refer to them, for
simplicity, by the names B1, B2, B3, are presented in the insets of Figs.
\ref{B1_S11_jumps},\ \ref{S11_4.39}, \ref{S11 two jumps} respectively. The
width of the feedlines and the thin part of the resonators was set to $0.4%
%TCIMACRO{\unit{mm}}%
%BeginExpansion
\mathrm{mm}%
%EndExpansion
$ to obtain characteristic impedance of $50%
%TCIMACRO{\unit{\U{3a9}}}%
%BeginExpansion
\mathrm{\Omega }%
%EndExpansion
$. The gap between the feedline and the resonators was set to $0.4%
%TCIMACRO{\unit{mm}}%
%BeginExpansion
\mathrm{mm}%
%EndExpansion
$ in the B1, B3 cases, and to $0.5%
%TCIMACRO{\unit{mm}}%
%BeginExpansion
\mathrm{mm}%
%EndExpansion
$ in B2 case. The frequency modes of B1, B2, B3 resonators were theoretically
calculated using a simple transmission line model, presented in appendix A,
and were also experimentally measured using vector network analyzer (NA). The
theoretical calculation was generally found to be in good agreement with the
measurement results, as discussed in Appendix A.

\section{Fabrication Process}

\textbf{\ }The sputtering of the NbN films was done using a dc-magnetron
sputtering system. All of the resonators reported here were deposited near
room temperature \cite{crystal NbN ambient temp},\cite{properties N2 and
normal resistivity},\cite{high Tc sc NbN room temp}, where no external heating
was applied. The system was usually pumped down prior to sputtering to
$3-8\cdot10^{-8}%
%TCIMACRO{\unit{torr}}%
%BeginExpansion
\mathrm{torr}%
%EndExpansion
$ base pressure (achieved overnight). The sputtering was done in Ar/N$_{2}$
atmosphere under current stabilization condition \cite{The bandwidth of Heb}.
The relative flow ratio of the two gases into the chamber and the total
pressure of the mixture were controlled by mass flow meters. The sputtering
usually started with a two minute pre-sputtering in the selected ambient
before removal of the shutter and deposition on the substrate. The sputtering
parameters of the three resonators are summarized in table 1. Following the
NbN deposition, the resonator features were patterned using standard
photolithography process, whereas the NbN etching was done using Ar ion-milling.

The fabricated resonators were characterized by relatively low $T_{c}$ for NbN
and relatively high normal resistivity $\rho,$ that is in good agreement with
Ref. \cite{properties N2 and normal resistivity}. $T_{c}$ measured for B1, B2
and B3 was $10.7,$ $6.8,$ $8.9$ $\left[
%TCIMACRO{\unit{K}}%
%BeginExpansion
\mathrm{K}%
%EndExpansion
\right]  $ respectively, whereas $\rho$ measured for B2 and B3 was $348$ and
$500$ $\left[  \mu%
%TCIMACRO{\unit{\U{3a9}}}%
%BeginExpansion
\mathrm{\Omega }%
%EndExpansion%
%TCIMACRO{\unit{cm}}%
%BeginExpansion
\mathrm{cm}%
%EndExpansion
\right]  $ respectively.

\begin{figure}
[ptb]
\begin{center}
\includegraphics[
height=2.885in,
width=3.2214in
]%
{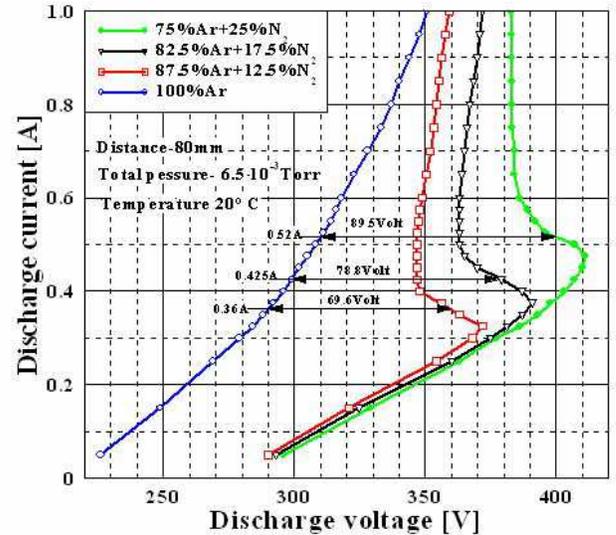}%
\caption{Discharge current vs. discharge voltage of the sputtering system
displaying current-voltage knee for different percentages of Argon and
Nitrogen, at ambient temperature of 20$\mathrm{{}^{\circ}{\rm C}}.$}%
\label{knee_temp}%
\end{center}
\end{figure}

To obtain resonators with reproducible physical properties we have used the
sputtering method discussed in \cite{The bandwidth of Heb}, where it was
claimed that reproducible parameters of films are assured, by keeping the
difference between the discharge voltage in a gas mixture, and in pure argon,
constant, for the same discharge current. In Fig. \ref{knee_temp} we show one
of the characterization measurements applied to our dc-magnetron sputtering
system, exhibiting a knee-shape graph of discharge current as a function of
discharge voltage. The knee-shape graph was obtained for different Ar/N$_{2}$
mixtures at room temperature as shown in the figure. The discharge voltage
difference measured in the presence of N$_{2}$ gas relative to the value
measured in pure argon at the same discharge current, is also pointed out in
the figure, corresponding to different currents and N$_{2}$ percentages.%

%TCIMACRO{\FRAME{ftbpFU}{3.2214in}{2.885in}{0pt}{\Qcb{Discharge current vs.
%discharge voltage of the sputtering system displaying current-voltage knee for
%different percentages of Argon and Nitrogen, at ambient temperature of
%20$\unit{\U{2103}}.$}}{\Qlb{knee_temp}}{process.eps}%
%width 3.2214in;  height 2.885in;  depth 0pt;  original-width 6.6642in;
%original-height 5.9655in;  cropleft "0";  croptop "1";  cropright "1";
%cropbottom "0";  filename '../process.eps';file-properties "XNPEU";}} }%
%BeginExpansion
%EndExpansion

\section{Measurement Results}

\bigskip All measurements presented in this paper have been conducted at
liquid helium temperature.

\subsection{$S_{11}$ Measurements}

\begin{figure}
[ptb]
\begin{center}
\includegraphics[
height=2.6013in,
width=3.1644in
]%
{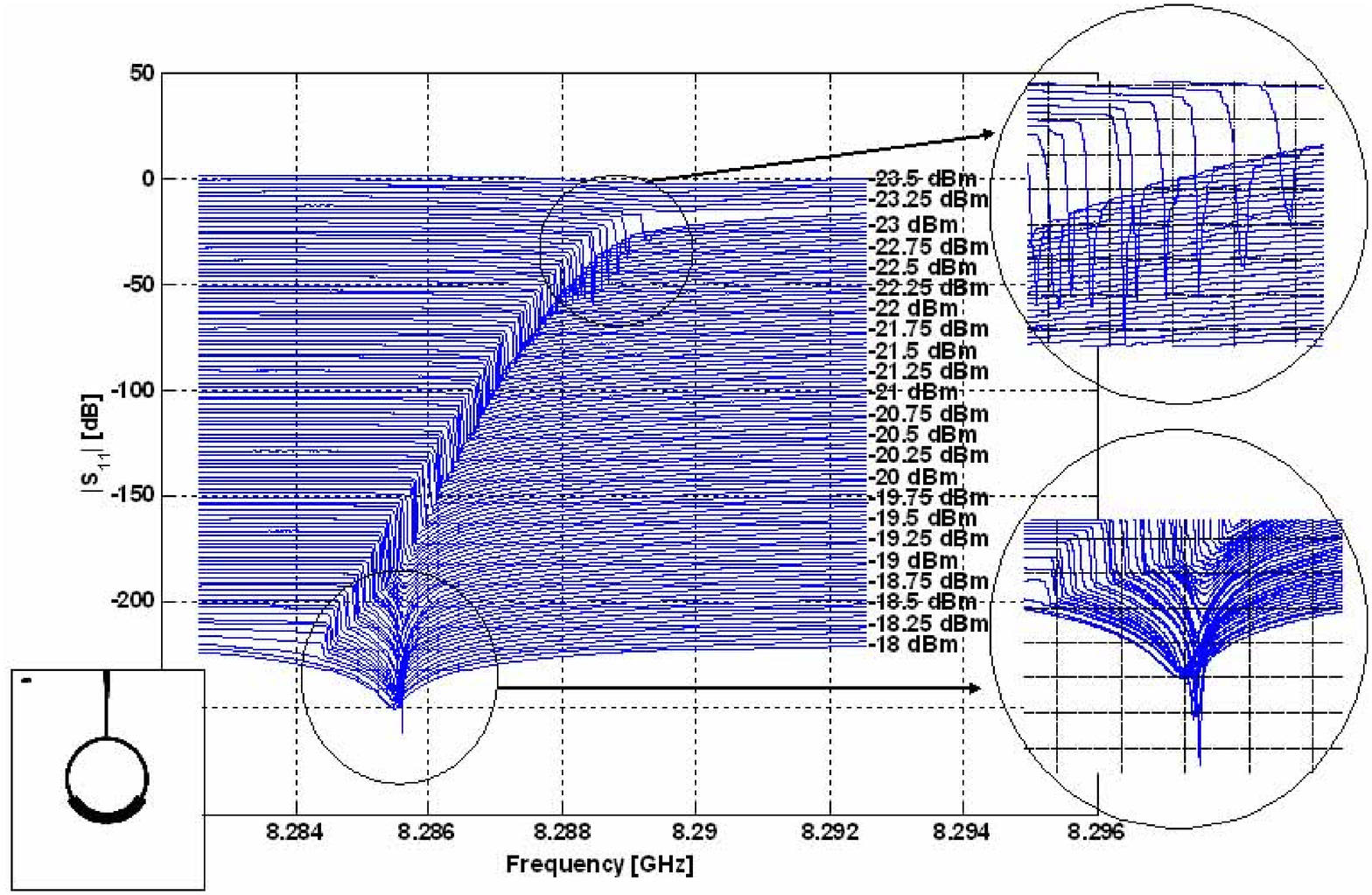}%
\caption{ $S_{11\text{ }}$measurement of B1 resonance at $\sim
8.26\mathrm{GHz}$ with 10$\mathrm{MHz}$ span, at low input powers,
exhibiting extraordinary nonlinear effects. The resonance curves corresponding
to different input powers were shifted by a constant offset for clarity. The
inset shows the resonator geometry.}%
\label{B1_S11_jumps}%
\end{center}
\end{figure}

The resonance response of the resonators was measured using the reflection
parameter $S_{11}$of a vector NA. The resonance response obtained for the
third mode of B1 resonator $\sim$8.26$%
%TCIMACRO{\unit{GHz}}%
%BeginExpansion
\mathrm{GHz}%
%EndExpansion
$ at low input powers, between $-23$ dBm and $-18$ dBm in steps of $0.05$
dBm$,$ is shown in Fig. \ref{B1_S11_jumps}. A small offset was applied between
the sequential graphs, corresponding to different input powers, for clarity,
and to emphasize the nonlinear evolution of the resonance response as the
input power is increased. The interesting characteristics of this nonlinear
evolution could be summarized as follows:%

%TCIMACRO{\FRAME{ftbpFU}{3.1644in}{2.6013in}{0pt}{\Qcb{ $S_{11\text{ }}%
%$measurement of B1 resonance at $\sim8.26\unit{GHz}$ with 10$\unit{MHz}$ span,
%at low input powers, exhibiting extraordinary nonlinear effects. The resonance
%curves corresponding to different input powers were shifted by a constant
%offset for clarity. The inset shows the resonator geometry.}}%
%{\Qlb{B1_S11_jumps}}{b1res.eps}{\special{ language "Scientific Word";
%type "GRAPHIC";  display "USEDEF";  valid_file "F";  width 3.1644in;
%height 2.6013in;  depth 0pt;  original-width 11.1301in;
%original-height 7.4088in;  cropleft "0";  croptop "1";  cropright "1";
%cropbottom "0";  filename '../B1res.eps';file-properties "XNPEU";}} }%
%BeginExpansion
%EndExpansion

1) In the power range between $-23.5$ dBm and $-23.25$ dBm, the resonance is
symmetrical and broad.

2) At input power level of $-23.25$ dBm, a sudden jump of about $-15$ dB
occurs in the resonance curve at the minima where the slope of the resonance
response is small.

3) As the input power is increased in steps of $0.05$ dBm the resonance
becomes asymmetrical, and the left jump shifts towards the lower frequencies gradually.

4) As we continue to increase the input power, the jumps decrease their height
but the resonance curve following the jumps becomes more symmetrical and
deeper, and at certain input power level we even witness a critical coupling
phenomenon where $S_{11}\left(  \omega\right)  $ at resonance is almost zero,
no power reflection is present.

5) The resonance becomes symmetrical again and broader and the bifurcations disappear.

6) All previously listed effects occur within a frequency span of 10$%
%TCIMACRO{\unit{MHz}}%
%BeginExpansion
\mathrm{MHz}%
%EndExpansion
$, power range of about 5 dBm, and power step of $0.05$ dBm.

Moreover, in order to estimate how narrow this resonance could be in the
vicinity of critical coupling, a separate $S_{11}$ measurement has been
applied near critical coupling power, using 1601 measurement points and
frequency span of 2 $%
%TCIMACRO{\unit{MHz}}%
%BeginExpansion
\mathrm{MHz}%
%EndExpansion
$. The bandwidth of the resonance curves measured, according to the +3 dB
method, was about $0.25\cdot10^{-5}%
%TCIMACRO{\unit{GHz}}%
%BeginExpansion
\mathrm{GHz}%
%EndExpansion
,$ whereas the ratio $f/\Delta f\sim3.3\cdot10^{6}.$\ 

\begin{figure}
[ptb]
\begin{center}
\includegraphics[
height=2.4976in,
width=3.2301in
]%
{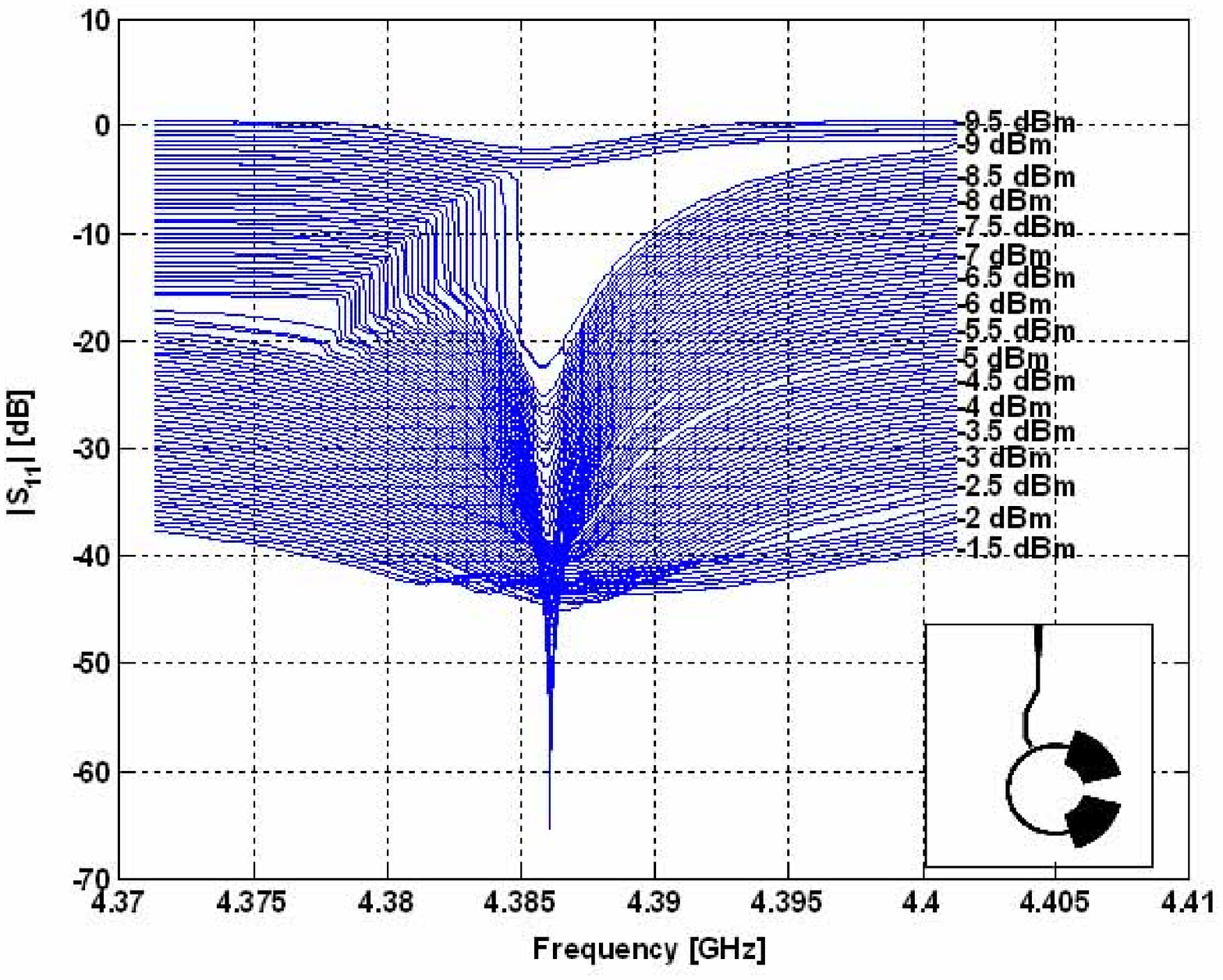}%
\caption{ $S_{11\text{ }}$measurement of B2 resonance at $\sim$
$4.385\mathrm{GHz}$ with 30$\mathrm{MHz}$ span, at low input
powers, exhibiting strong nonlinear effects. The resonance curves
corresponding to different input powers were shifted by a constant offset for
clarity. The inset shows the resonator geometry.}%
\label{S11_4.39}%
\end{center}
\end{figure}

\begin{figure}
[ptb]
\begin{center}
\includegraphics[
height=2.4232in,
width=3.2119in
]%
{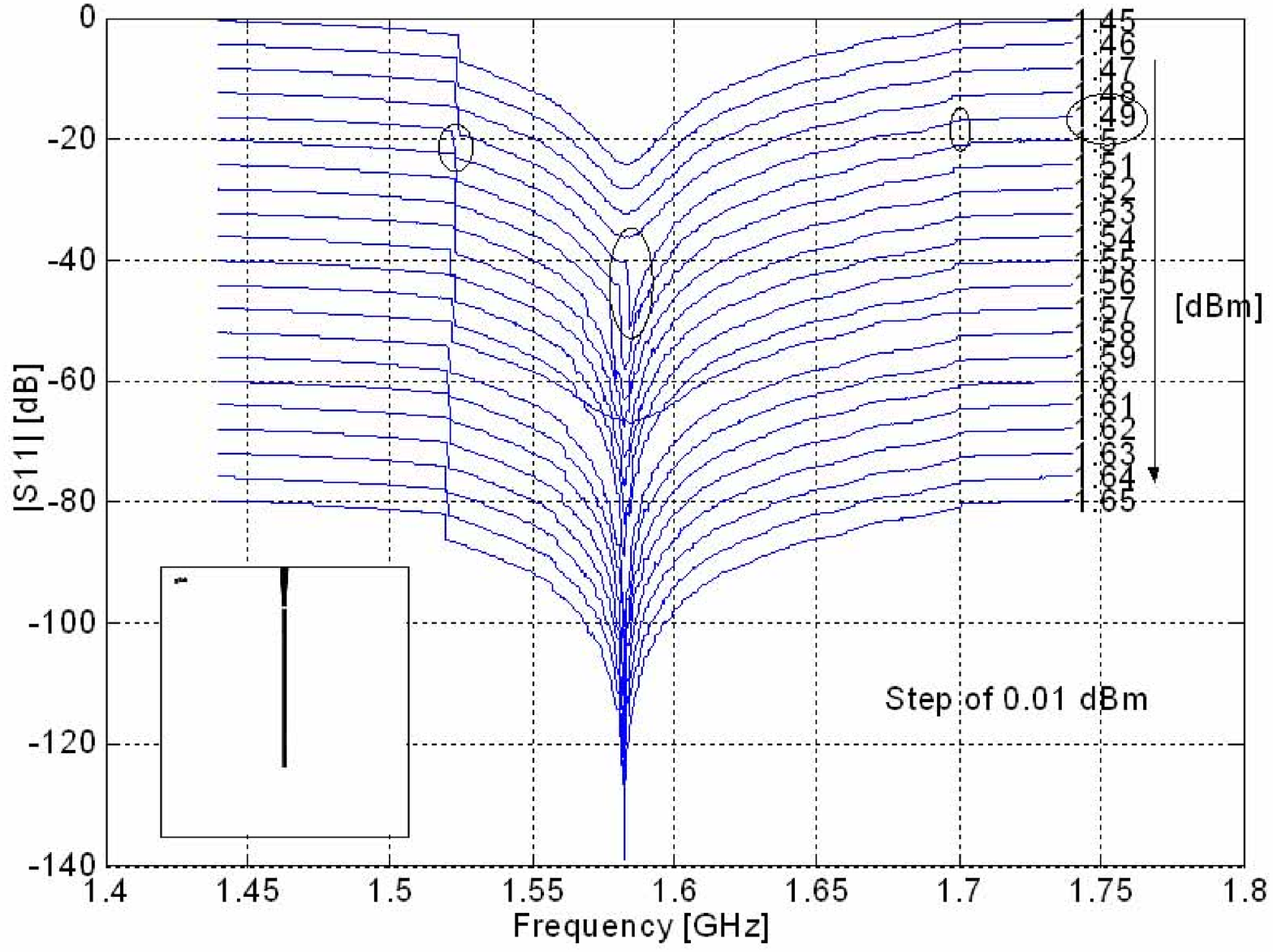}%
\caption{Nonlinear response of B3 resonance at $\sim1.6$ $\mathrm{GHz},$
corresponding to input power levels increasing in 0.01 dBm. At input power of
1.49 dBm we observe two obvious bifurcations in the resonance band, and
another small one at the right side, marked with circles. The different
resonance curves were shifted by a constant offset for clarity. The inset
shows the resonator geometry.}%
\label{S11 two jumps}%
\end{center}
\end{figure}

Similar behavior to that exhibited by the nonlinear third mode of resonator B1
can be clearly seen in Fig. \ref{S11_4.39} and Fig. \ref{S11 two jumps}, which
show the nonlinear dynamic evolution of the second mode of resonator B2 and
the first of B3 respectively. The main differences between the figures are:

1) The power levels at which these nonlinear effects take place. Whereas in B1
case they happen between $-23$ dBm and $-18$ dBm, in B2 case they happen
between $-9.5$ dBm and $-1.5$ dBm and in B3 case they happen around $\ 1$ dBm.

2) In Fig. \ref{S11 two jumps} corresponding to B3 resonator we witness two
apparent bifurcations within the resonance band as indicated by circles on the
figure, a feature that we did not encounter in Figs. \ref{B1_S11_jumps} and
\ref{S11_4.39}.%

%TCIMACRO{\FRAME{ftbpFU}{3.2301in}{2.4976in}{0pt}{\Qcb{ $S_{11\text{ }}%
%$measurement of B2 resonance at $\sim$ $4.385\unit{GHz}$ with 30$\unit{MHz}$
%span, at low input powers, exhibiting strong nonlinear effects. The resonance
%curves corresponding to different input powers were shifted by a constant
%offset for clarity. The inset shows the resonator geometry.}}{\Qlb{S11_4.39}%
%}{b2res.eps}{\special{ language "Scientific Word";  type "GRAPHIC";
%display "USEDEF";  valid_file "F";  width 3.2301in;  height 2.4976in;
%depth 0pt;  original-width 9.5008in;  original-height 7.2428in;
%cropleft "0";  croptop "1";  cropright "1";  cropbottom "0";
%filename '../B2res.eps';file-properties "XNPEU";}} }%
%BeginExpansion
%EndExpansion
%

%TCIMACRO{\FRAME{ftbpFU}{3.2119in}{2.4232in}{0pt}{\Qcb{Nonlinear response of B3
%resonance at $\sim1.6$ $\unit{GHz},$ corresponding to input power levels
%increasing in 0.01 dBm. At input power of 1.49 dBm we observe two obvious
%bifurcations in the resonance band, and another small one at the right side,
%marked with circles. The different resonance curves were shifted by a constant
%offset for clarity. The inset shows the resonator geometry.}}{\Qlb{S11 two
%jumps}}{b3res.eps}{\special{ language "Scientific Word";  type "GRAPHIC";
%display "USEDEF";  valid_file "F";  width 3.2119in;  height 2.4232in;
%depth 0pt;  original-width 12.794in;  original-height 9.0477in;
%cropleft "0";  croptop "1";  cropright "1";  cropbottom "0";
%filename 'B3res.eps';file-properties "XNPEU";}} }%
%BeginExpansion
%EndExpansion

\subsection{Verifications}

\begin{figure}
[ptb]
\begin{center}
\includegraphics[
height=2.2009in,
width=3.1808in
]%
{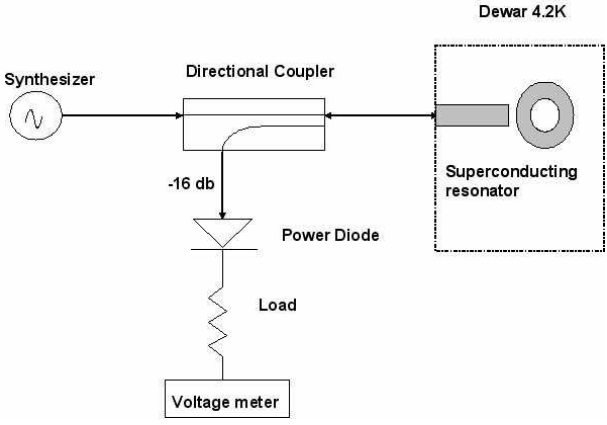}%
\caption{A setup that was used to verify the occurrence of the bifuractions at
the nonlinear resonance (B1 third mode), measured previously by NA.}%
\label{verify_diode}%
\end{center}
\end{figure}

\begin{figure}
[ptb]
\begin{center}
\includegraphics[
height=2.2943in,
width=3.1912in
]%
{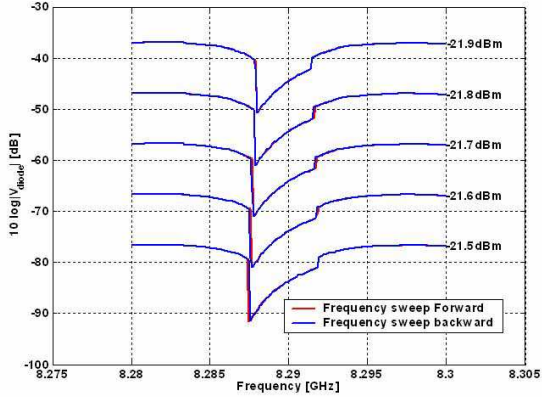}%
\caption{Nonlinear response of B1 third mode measured using the configuration
shown in Fig. \ref{verify_diode}. The red line represents a forward scan
whereas the blue line represents a backward scan. Two abrupt jumps appear at
both sides of the resonance curve and small hysteresis loops are present at
the vicinity of the jumps. The resonance curves corresponding to different
input powers were shifted by a constant offset for clarity.}%
\label{verify_diode_meas}%
\end{center}
\end{figure}

In order to verify that the bifurcation feature, previously measured using NA
$S_{11}$ parameter, is not a measurement artifact, we applied a different
measurement configuration, shown in Fig. \ref{verify_diode}, where we scanned
the frequency axis with CW mode of an Agilent synthesizer and measured the
reflected power from the resonator by a power diode and voltage meter. The
load that appears in Fig. \ref{verify_diode} following the diode is an Agilent
load used in order to extend the linear regime of the power diode. The results
of this measurement configuration are shown in Fig. \ref{verify_diode_meas}.
The frequency scan around the resonance was done using 201 points in each
direction (forward and backward). A small hysteresis loop can be seen around
the two bifurcations.%

%TCIMACRO{\FRAME{ftbpFU}{3.1808in}{2.2009in}{0pt}{\Qcb{A setup that was used to
%verify the occurrence of the bifuractions at the nonlinear resonance (B1 third
%mode), measured previously by NA.}}{\Qlb{verify_diode}}{diode.eps}%
%width 3.1808in;  height 2.2009in;  depth 0pt;  original-width 8.3636in;
%original-height 5.7718in;  cropleft "0";  croptop "1";  cropright "1";
%cropbottom "0";  filename '../diode.eps';file-properties "XNPEU";}} }%
%BeginExpansion
%EndExpansion
%

%TCIMACRO{\FRAME{ftbpFU}{3.1912in}{2.2943in}{0pt}{\Qcb{Nonlinear response of B1
%third mode measured using the configuration shown in Fig. \ref{verify_diode}.
%The red line represents a forward scan whereas the blue line represents a
%backward scan. Two abrupt jumps appear at both sides of the resonance curve
%and small hysteresis loops are present at the vicinity of the jumps. The
%resonance curves corresponding to different input powers were shifted by a
%constant offset for clarity.}}{\Qlb{verify_diode_meas}}{dioderesults.eps}%
%width 3.1912in;  height 2.2943in;  depth 0pt;  original-width 10.0024in;
%original-height 7.1728in;  cropleft "0";  croptop "1";  cropright "1";
%cropbottom "0";  filename '../dioderesults.eps';file-properties "XNPEU";}} }%
%BeginExpansion
%EndExpansion

\subsection{Abrupt Bifurcations}

\begin{figure}
[ptb]
\begin{center}
\includegraphics[
height=2.239in,
width=2.84in
]%
{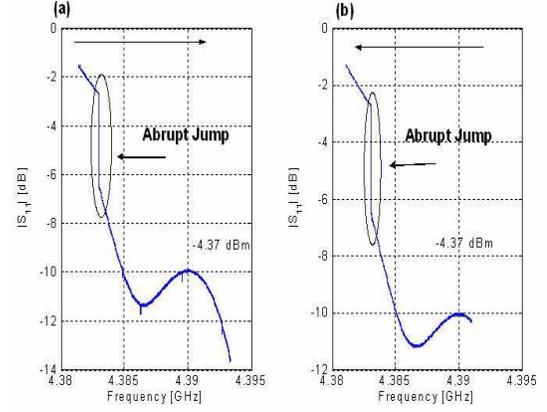}%
\caption{(a) A forward CW mode scan using NA, in a 12$\mathrm{MHz}$ span
around the left jump of the nonlinear resonance of B2. The scan includes
20,000 frequencies, which is equivalent to a frequency step of $\sim
$600$\mathrm{Hz}$ between the data points. In spite of this small
frequency step, the jump still occurs between just two sequential frequencies.
(b) A backward CW mode scan using NA, in a 10$\mathrm{MHz}$ span around
the left jump of the nonlinear resonance of B2. The scan includes 4000
frequencies, which is equivalent to a frequency step of $\sim$%
2.5$\mathrm{kHz}$ between the data points. Also for this case the jump
occurs between just two sequential frequencies.}%
\label{8001pointcw}%
\end{center}
\end{figure}

In attempt to find out whether the resonance curve of these nonlinear
resonances changes its form along two or more frequency points, further
measurements where carried out using NA, where we scanned the frequency axis
in the vicinity of the jump with high frequency resolution. The measurement
results corresponding to frequency step of $\sim$600 $%
%TCIMACRO{\unit{Hz}}%
%BeginExpansion
\mathrm{Hz}%
%EndExpansion
$ and $\sim$2.5 $%
%TCIMACRO{\unit{kHz}}%
%BeginExpansion
\mathrm{kHz}%
%EndExpansion
$ are presented in Fig. \ref{8001pointcw} plot (a) and (b) respectively,
indicating abrupt transition between two bistable states.%

%TCIMACRO{\FRAME{ftbpFU}{2.84in}{2.239in}{0pt}{\Qcb{(a) A forward CW mode scan
%using NA, in a 12$\unit{MHz}$ span around the left jump of the nonlinear
%resonance of B2. The scan includes 20,000 frequencies, which is equivalent to
%a frequency step of $\sim$600$\unit{Hz}$ between the data points. In spite of
%this small frequency step, the jump still occurs between just two sequential
%frequencies. (b) A backward CW mode scan using NA, in a 10$\unit{MHz}$ span
%around the left jump of the nonlinear resonance of B2. The scan includes 4000
%frequencies, which is equivalent to a frequency step of $\sim$2.5$\unit{kHz}$
%between the data points. Also for this case the jump occurs between just two
%sequential frequencies.}}{\Qlb{8001pointcw}}{highfreqres.eps}%
%{\special{ language "Scientific Word";  type "GRAPHIC";  display "USEDEF";
%valid_file "F";  width 2.84in;  height 2.239in;  depth 0pt;
%original-width 9.6816in;  original-height 6.0027in;  cropleft "0";
%filename '../highfreqres.eps';file-properties "XNPEU";}} }%
%BeginExpansion
%EndExpansion

\subsection{Frequency Hysteresis}

\begin{figure}
[ptb]
\begin{center}
\includegraphics[
height=2.4353in,
width=3.4143in
]%
{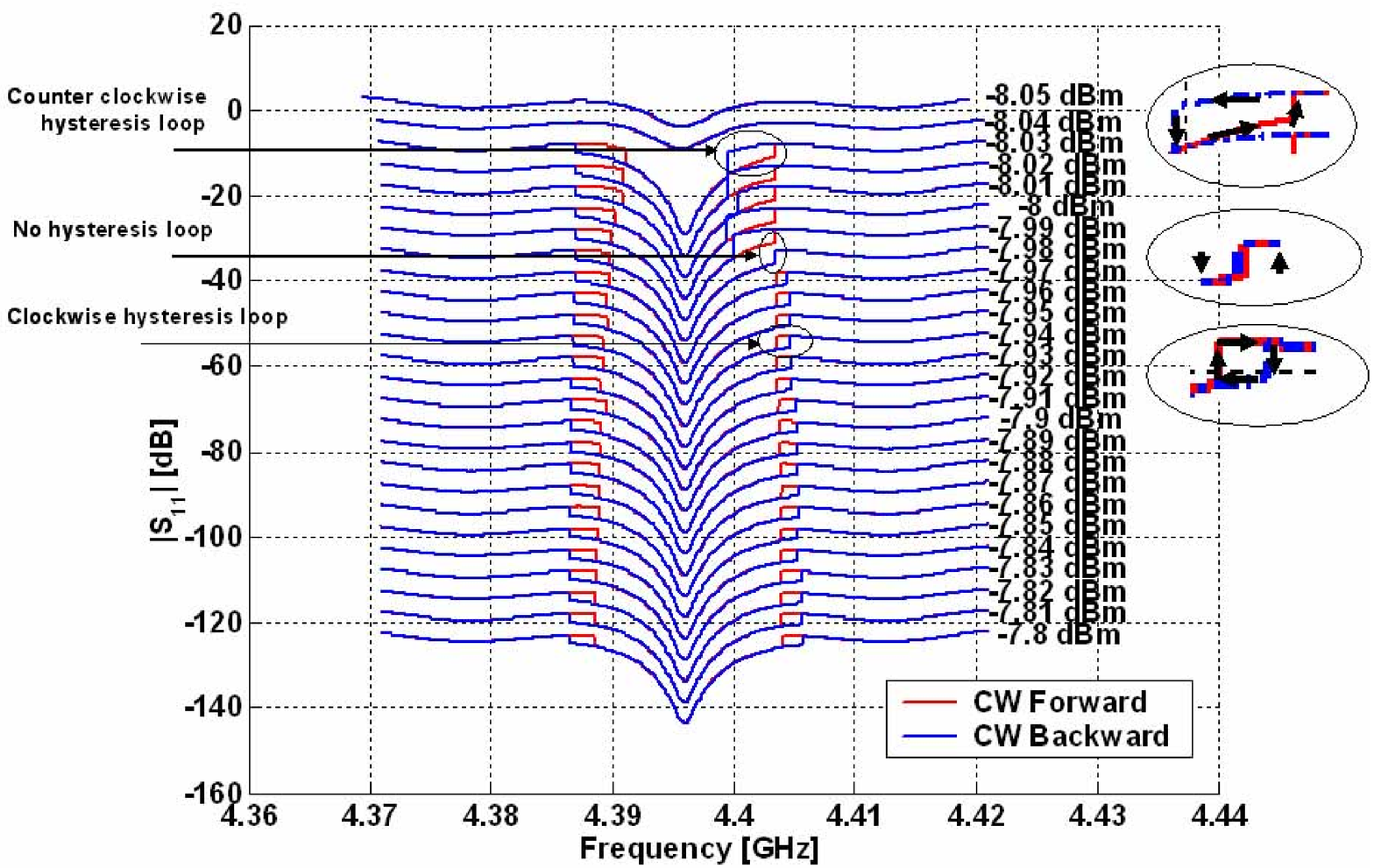}%
\caption{Forward and backward scan measurement, performed using NA, measuring
B2 second mode nonlinear resonance. The red line represents a forward scan
whereas the blue line represents a backward scan. The graphs exhibit clear
hysteresis loops forming at the vicinity of the bifurcations, and hysteresis
loop changing direction as the input power is increased. The resonance curves
corresponding to different input powers were shifted by a constant offset for
clarity.}%
\label{hysteresis_behavior_B2}%
\end{center}
\end{figure}

Applying forward and backward frequency sweeps to these resonators, reveals a
very interesting hysteretic behavior. In Fig. \ref{hysteresis_behavior_B2} we
show a representative frequency scan of B2 second mode, applied in both
directions, featuring the following nonlinear dynamic behavior:

1) At low input powers $-8.05$ dBm and $-8.04$ dBm, the resonance is
symmetrical and there is no hysteresis.

2) As the power is increased by 0.01 dBm to $-8.03$ dBm, two bifurcations
occur at both sides of the resonance response and hysteresis loops \ form at
the bistable regions.

3) As we continue to increase the input power gradually, the hysteresis loop,
associated with the right bifurcation, changes direction. At first it
circulates counterclockwise between $-8.03$ dBm and $-7.99$ dBm, at $-7.98$
dBm the two opposed bifurcations, at the right side, meet and no hysteresis is
detected, as we increase the power further the right hysteresis loop appears
again, circulating, this time, in the opposite direction, clockwise.

Furthermore, it is worth mentioning that the hysteresis loops changing
direction are not unique to this resonator, or to the bifurcation occurring on
the right side of the resonance. It appears also in the modes of B1, and it
occurs at the left side bifurcation as well, but at different power level.%

%TCIMACRO{\FRAME{ftbpFU}{3.4143in}{2.4353in}{0pt}{\Qcb{Forward and backward
%scan measurement, performed using NA, measuring B2 second mode nonlinear
%resonance. The red line represents a forward scan whereas the blue line
%represents a backward scan. The graphs exhibit clear hysteresis loops forming
%at the vicinity of the bifurcations, and hysteresis loop changing direction as
%the input power is increased. The resonance curves corresponding to different
%input powers were shifted by a constant offset for clarity.}}%
%{\Qlb{hysteresis_behavior_B2}}{hysteresis.eps}%
%{\special{ language "Scientific Word";  type "GRAPHIC";  display "USEDEF";
%valid_file "F";  width 3.4143in;  height 2.4353in;  depth 0pt;
%original-width 13.2256in;  original-height 8.8669in;  cropleft "0";
%filename '../hysteresis.eps';file-properties "XNPEU";}} }%
%BeginExpansion
%EndExpansion

\subsection{Multiple Bifurcations}

\begin{figure}
[ptb]
\begin{center}
\includegraphics[
height=2.3627in,
width=2.9058in
]%
{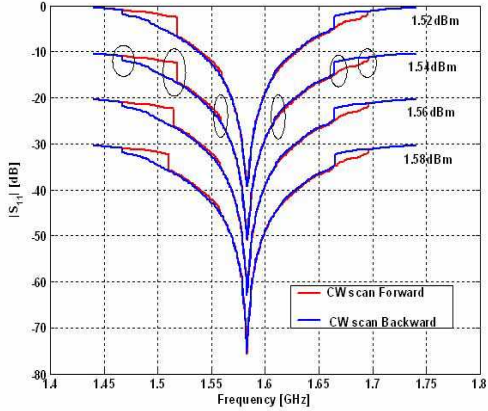}%
\caption{$S_{11}$ parameter measurement of the first resonance of B3 at input
powers 1.52 dBm through 1.58 dBm in steps of 0.02 dBm using forward and
backward CW mode scan of NA employing 2000 measurement points in each
direction. The graph shows clearly 4 jumps within the band of the resonance in
each direction. The resonance curves corresponding to different input powers
were shifted by a constant offset for clarity.}%
\label{2000point_four_jump}%
\end{center}
\end{figure}

Frequency sweep applied to B3 first resonance in both directions, exhibits yet
another feature, in addition to the two bifurcations at the sides of the
resonance curve, which we have seen earlier, there are another two smaller
bifurcations accompanied with hysteresis within the resonance lineshape,
adding up to 4 bifurcations in each scan direction, as can be seen in Fig.
\ref{2000point_four_jump}. This feature may have a special significance in
explaining the physical origin of these nonlinearities.%

%TCIMACRO{\FRAME{ftbpFU}{2.9058in}{2.3627in}{0pt}{\Qcb{$S_{11}$ parameter
%measurement of the first resonance of B3 at input powers 1.52 dBm through 1.58
%dBm in steps of 0.02 dBm using forward and backward CW mode scan of NA
%employing 2000 measurement points in each direction. The graph shows clearly 4
%jumps within the band of the resonance in each direction. The resonance curves
%corresponding to different input powers were shifted by a constant offset for
%clarity.}}{\Qlb{2000point_four_jump}}{b3hys.eps}%
%{\special{ language "Scientific Word";  type "GRAPHIC";  display "USEDEF";
%valid_file "F";  width 2.9058in;  height 2.3627in;  depth 0pt;
%original-width 9.0001in;  original-height 6.5034in;  cropleft "0";
%filename '../B3hys.eps';file-properties "XNPEU";}} }%
%BeginExpansion
%EndExpansion

\section{Discussion}

This unusual dynamic behavior of our NbN \ films, demonstrated earlier, highly
suggest built-in Josephson junctions, forming at the boundaries of the
granular NbN columnar structure \cite{rf superconducting properties of
reactively sputtered NbN},\cite{superconducting properties and NbN structure},
as the underlying physical mechanism responsible for the observed
nonlinearities \cite{rf residual losses Halbritter}. Another physical
mechanism that may be considered as a strong candidate for explaining the
effects, is the local heating mechanism which was hypothesized as the source
of notches and switching effects observed in HTS films \cite{HTS patch
antenna},\cite{power dependent effects observed for sc stripline
resonator},\cite{mw power handling weak links thermal effects}. Nevertheless
recent measurements done on SQUID ring containing a single Josephson junction
inductively coupled to a radio frequency resonant circuit \cite{opposed
hammerhead}, \cite{nonlinear multilevel}, \cite{Pinch resonances in rf}
exhibiting opposed bifurcations, hammerhead resonance lineshape and similar
effects, suggest similar physical mechanism and further support the former hypothesis.

\section{Conclusion}

%if have a single appendix:
%or
%do not use \section anymore after \appendix, only \section*
%is possibly needed
In the course of this experimental work we have fabricated several stripline
NbN resonators dc-magnetron sputtered on sapphire substrates at room
temperature implementing different geometries. The resonators have exhibited
similar and unusual nonlinear effects in their resonance response curves. The
onset of the nonlinear effects in these NbN resonators varied between the
different resonators, but usually occurred at relatively low powers, typically
2-3 orders of magnitude lower than Nb for example. Among the nonlinear effects
observed: abrupt and multiple bifurcations in the resonance curve, power
dependent resonance frequency shift, hysteresis loops in the vicinity of the
bifurcations, hysteresis loops changing direction, and critical coupling
phenomenon. Weak links forming in the NbN films are hypothesized as the source
of the nonlinearities. Further study of these effects under other modes of
operation and measurement conditions would be carried in the future, in order
to substantiate our understanding of these extraordinary effects.

%use appendices with more than one appendix
%then use \section to start each appendix
%you must declare a \section before using any
%\subsection or using \label (\appendices by itself
%starts a section numbered zero.)
%Use this command to get the appendices' numbers in "A", "B" instead of the
%default capitalized Roman numerals ("I", "II", etc.).
%However, the capital letter form may result in awkward subsection numbers
%(such as "A-A"). Capitalized Roman numerals are the default.
%\useRomanappendicesfalse
\appendices

\section{Resonance frequency calculation of B1, B2, B3 resonators}

\begin{figure}
[ptb]
\begin{center}
\includegraphics[
height=1.5264in,
width=2.7121in
]%
{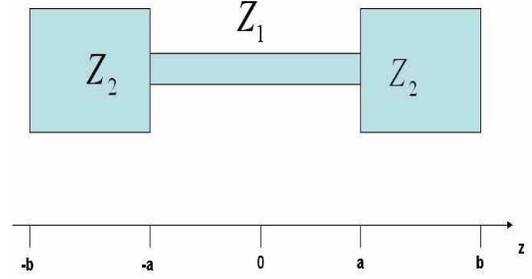}%
\caption{Resonator general geometry model.}%
\label{Resonator general model}%
\end{center}
\end{figure}

The calculation process of the resonance frequencies of B1 and B2 makes use of
opposite traveling voltage-current waves method \cite{planar microstrip ring
resonator filters,varacter-tuned ring circuits}. For this purpose we model B1
and B2 resonators as a straight transmission line extending in the z-direction
with two characteristic impedance regions $Z_{1}$ and $Z_{2}$ as shown in Fig.
\ref{Resonator general model}.%

%TCIMACRO{\FRAME{ftbpFU}{2.7121in}{1.5264in}{0pt}{\Qcb{Resonator general
%geometry model.}}{\Qlb{Resonator general model}}{model.eps}%
%{\special{ language "Scientific Word";  type "GRAPHIC";  display "USEDEF";
%valid_file "F";  width 2.7121in;  height 1.5264in;  depth 0pt;
%original-width 9.404in;  original-height 4.0534in;  cropleft "0";
%filename '../model.eps';file-properties "XNPEU";}} }%
%BeginExpansion
%EndExpansion

The equivalent voltage along the resonator transmission line would be given,
in general, by a standing waves expression in the form:%

\begin{equation}
V\left(  z\right)  =\left\{
\begin{array}
[c]{cc}%
A_{+}\cos\beta\left(  z-a\right)  +B_{+}\sin\beta\left(  z-a\right)  &
z\epsilon(a,b)\\
A\cos\beta z+B\sin\beta z & z\epsilon(-a,a)\\
A_{-}\cos\beta\left(  z+a\right)  +B_{-}\sin\beta\left(  z+a\right)  &
z\epsilon(-b,-a)
\end{array}
\right.
\end{equation}

where $\beta$ $\left(  =2\pi f\sqrt{\varepsilon_{r}}/c\right)  $ is the
propagation constant along the transmission line, and \ $A_{+},B_{+}%
,A,B,A_{-},B_{-}$ are constants that can be determined using boundary
conditions and applied power amplitude. However due to the symmetry of the
problem $z\longleftrightarrow-z,$ we expect the solutions to have defined
parity, where $V\left(  z\right)  =V\left(  -z\right)  $ for symmetric
solution and $V\left(  z\right)  =-V\left(  -z\right)  $ for antisymmetric
solution. Thus by taking advantage of this property and demanding that
$V\left(  z\right)  $ be continuous at $z=a$ and $z=-a$, one gets:%

\begin{tabular}
[c]{l}%
$V_{sym}\left(  z\right)  =\left\{
\begin{array}
[c]{cc}%
\begin{array}
[c]{c}%
\cos\beta a\cos\beta\left(  z-a\right) \\
+B_{s}\sin\beta\left(  z-a\right)
\end{array}
& z\epsilon(a,b)\\
\cos\beta z & z\epsilon(-a,a)\\%
\begin{array}
[c]{c}%
\cos\beta a\cos\beta\left(  z+a\right) \\
-B_{s}\sin\beta\left(  z+a\right)
\end{array}
& z\epsilon(-b,-a)
\end{array}
\right.  $\\
$V_{anti}\left(  z\right)  =\left\{
\begin{array}
[c]{cc}%
\begin{array}
[c]{c}%
\sin\beta a\cos\beta\left(  z-a\right) \\
+B_{a}\sin\beta\left(  z-a\right)
\end{array}
& z\epsilon(a,b)\\
\sin\beta z & z\epsilon(-a,a)\\%
\begin{array}
[c]{c}%
-\sin\beta a\cos\beta\left(  z+a\right) \\
+B_{a}\sin\beta\left(  z+a\right)
\end{array}
& z\epsilon(-b,-a)
\end{array}
\right.  $%
\end{tabular}

where $V_{sym}\left(  z\right)  $ stands for the symmetric solution whereas
$V_{anti}\left(  z\right)  $ for the antisymmetric solution. To calculate the
value of the new constants $B_{s},B_{a}$ , we require that the equivalent
current $I\left(  z\right)  $ along the transmission line, which is given by
$I\left(  z\right)  =(i/\beta Z_{i})dV/dz$ where $Z_{i}$ is the characteristic
impedance of the line, be continuous at $z=a$ and $z=-a.$ Following this
requirement one gets $B_{s}=-\eta\sin\left(  \beta a\right)  $ and $B_{a}%
=\eta\cos\left(  \beta a\right)  $, where $\eta=Z_{2}/Z_{1}$. The symmetric
and antisymmetric solutions of $V\left(  z\right)  $ and $I\left(  z\right)  $
are given by:%

\begin{tabular}
[c]{l}%
$V_{sym}\left(  z\right)  =\left\{
\begin{array}
[c]{cc}%
\begin{array}
[c]{c}%
\cos\beta a\cos\beta\left(  z-a\right) \\
-\eta\sin\beta a\sin\beta\left(  z-a\right)
\end{array}
& z\epsilon(a,b)\\
\cos\beta z & z\epsilon(-a,a)\\%
\begin{array}
[c]{c}%
\cos\beta a\cos\beta\left(  z+a\right) \\
+\eta\sin\beta a\sin\beta\left(  z+a\right)
\end{array}
& z\epsilon(-b,-a)
\end{array}
\right.  $\\
$V_{anti}\left(  z\right)  =\left\{
\begin{array}
[c]{cc}%
\begin{array}
[c]{c}%
\sin\beta a\cos\beta\left(  z-a\right) \\
+\eta\cos\beta a\sin\beta\left(  z-a\right)
\end{array}
& z\epsilon(a,b)\\
\sin\left(  \beta z\right)  & z\epsilon(-a,a)\\%
\begin{array}
[c]{c}%
-\sin\left(  \beta a\right)  \cos\beta\left(  z+a\right) \\
+\eta\cos\left(  \beta a\right)  \sin\beta\left(  z+a\right)
\end{array}
& z\epsilon(-b,-a)
\end{array}
\right.  $%
\end{tabular}

\begin{tabular}
[c]{l}%
$I_{sym}\left(  z\right)  =\left\{
\begin{array}
[c]{cc}%
\begin{array}
[c]{c}%
(\frac{-i\cos\beta a\sin\beta\left(  z-a\right)  }{Z_{2}}\\
-\frac{i\eta\sin\beta a\cos\beta\left(  z-a\right)  }{Z_{2}})
\end{array}
& z\epsilon(a,b)\\
-\frac{i\sin\left(  \beta z\right)  }{Z_{1}} & z\epsilon(-a,a)\\%
\begin{array}
[c]{c}%
(\frac{-i\cos\beta a\sin\beta\left(  z+a\right)  }{Z_{2}}\\
+\frac{i\eta\sin\beta a\cos\beta\left(  z+a\right)  }{Z_{2}})
\end{array}
& z\epsilon(-b,-a)
\end{array}
\right.  $\\
$I_{anti}\left(  z\right)  =\left\{
\begin{array}
[c]{cc}%
\begin{array}
[c]{c}%
(\frac{-i\sin\beta a\sin\beta\left(  z-a\right)  }{Z_{2}}\\
+\frac{i\eta\cos\beta a\cos\beta\left(  z-a\right)  }{Z_{2}})
\end{array}
& z\epsilon(a,b)\\
\frac{i\cos\beta z}{Z_{1}} & z\epsilon(-a,a)\\%
\begin{array}
[c]{c}%
(\frac{i\sin\beta a\sin\beta\left(  z+a\right)  }{Z_{2}}\\
+\frac{i\eta\cos\beta a\cos\beta\left(  z+a\right)  }{Z_{2}})
\end{array}
& z\epsilon(-b,-a)
\end{array}
\right.  $%
\end{tabular}

\subsection{B1 Resonator}

Since the resonator ends are shorted we demand $V\left(  b\right)  =V\left(
-b\right)  .$

The symmetric case:

In this case we either have maximum or minimum at $V\left(  b\right)  $ thus
we get $I\left(  b\right)  =0,$ yielding the following condition on the
resonance frequencies$:$%

\begin{equation}
\cos\left(  \beta a\right)  \sin\left(  \beta\left(  b-a\right)  \right)
+\eta\sin\left(  \beta a\right)  \cos\left(  \beta\left(  b-a\right)  \right)
=0
\end{equation}

The antisymmetric case:

From the antisymmetric $V\left(  -b\right)  =-V\left(  -b\right)  $ and the
continuity $V\left(  b\right)  =V\left(  -b\right)  $ conditions, we get
$\ V\left(  b\right)  =V\left(  -b\right)  =0,$ which yields:%

\begin{equation}
\sin\left(  \beta a\right)  \cos\left(  \beta\left(  b-a\right)  \right)
+\eta\cos\left(  \beta a\right)  \sin\left(  \beta\left(  b-a\right)  \right)
=0
\end{equation}

Substituting the following numerical values $\eta=Z_{2}/Z_{1}=0.5,l_{1}=a=13%
%TCIMACRO{\unit{mm}}%
%BeginExpansion
\mathrm{mm}%
%EndExpansion
,l_{2}=b-a=6.5%
%TCIMACRO{\unit{mm}}%
%BeginExpansion
\mathrm{mm}%
%EndExpansion
$ into the above resonance frequency conditions and solving for frequencies
below $10%
%TCIMACRO{\unit{GHz}}%
%BeginExpansion
\mathrm{GHz}%
%EndExpansion
$, yields the following solutions ($2.5035%
%TCIMACRO{\unit{GHz}}%
%BeginExpansion
\mathrm{GHz}%
%EndExpansion
,$ $5.697%
%TCIMACRO{\unit{GHz}}%
%BeginExpansion
\mathrm{GHz}%
%EndExpansion
,$ $8.1647%
%TCIMACRO{\unit{GHz} }%
%BeginExpansion
\mathrm{GHz}
%EndExpansion
$) to the symmetric case, and ($2.9804%
%TCIMACRO{\unit{GHz}}%
%BeginExpansion
\mathrm{GHz}%
%EndExpansion
,$ $5.1786%
%TCIMACRO{\unit{GHz}}%
%BeginExpansion
\mathrm{GHz}%
%EndExpansion
,$ $8.1647%
%TCIMACRO{\unit{GHz}}%
%BeginExpansion
\mathrm{GHz}%
%EndExpansion
$)\ to the antisymmetric case, with doubly degenerate mode at $8.1647%
%TCIMACRO{\unit{GHz}}%
%BeginExpansion
\mathrm{GHz}%
%EndExpansion
$. By comparing these calculated resonances to the directly measured
resonances of B1 resonator, obtained using a broadband $S_{11}$ measurement
($2.5812%
%TCIMACRO{\unit{GHz}}%
%BeginExpansion
\mathrm{GHz}%
%EndExpansion
,5.6304%
%TCIMACRO{\unit{GHz}}%
%BeginExpansion
\mathrm{GHz}%
%EndExpansion
,8.4188%
%TCIMACRO{\unit{GHz}}%
%BeginExpansion
\mathrm{GHz}%
%EndExpansion
$)$,$ we find that the excited resonances correspond to the symmetrical case
only. The antisymmetric modes do not get excited because they have a voltage
node at the feeding line position.

\subsection{B2 Resonator}

Since the resonator ends are open-circuited we demand $I\left(  b\right)
=I(-b)=0.$

The symmetric case:

We require that the current associated with the symmetric voltage, vanishes:%

\begin{equation}
\cos\left(  \beta a\right)  \sin\left[  \beta\left(  b-a\right)  \right]
+\eta\sin\left(  \beta a\right)  \cos\left[  \beta\left(  b-a\right)  \right]
=0
\end{equation}

The antisymmetric case:

We require that the current associated with the antisymmetric voltage, vanishes:%

\begin{equation}
-\sin\left(  \beta a\right)  \sin\left(  \beta\left(  b-a\right)  \right)
+\eta\cos\left(  \beta a\right)  \cos\beta\left(  b-a\right)  =0
\end{equation}

Substituting the following numerical values $\eta=Z_{2}/Z_{1}%
=49.9/10.4=\allowbreak4.\,\allowbreak79,l_{1}=a=11.97%
%TCIMACRO{\unit{mm}}%
%BeginExpansion
\mathrm{mm}%
%EndExpansion
,l_{2}=b-a=6.43%
%TCIMACRO{\unit{mm}}%
%BeginExpansion
\mathrm{mm}%
%EndExpansion
$ into the above resonance frequency conditions and solving for frequencies
below $10%
%TCIMACRO{\unit{GHz}}%
%BeginExpansion
\mathrm{GHz}%
%EndExpansion
$, yields the following solutions ($2.6486%
%TCIMACRO{\unit{GHz}}%
%BeginExpansion
\mathrm{GHz}%
%EndExpansion
,$ $6.0288%
%TCIMACRO{\unit{GHz}}%
%BeginExpansion
\mathrm{GHz}%
%EndExpansion
,$ $8.5588%
%TCIMACRO{\unit{GHz}}%
%BeginExpansion
\mathrm{GHz}%
%EndExpansion
$) to the symmetric case, and ($1.1763%
%TCIMACRO{\unit{GHz}}%
%BeginExpansion
\mathrm{GHz}%
%EndExpansion
,$ $4.3698%
%TCIMACRO{\unit{GHz}}%
%BeginExpansion
\mathrm{GHz}%
%EndExpansion
,$ $7.4597%
%TCIMACRO{\unit{GHz}}%
%BeginExpansion
\mathrm{GHz}%
%EndExpansion
,$ $9.7778%
%TCIMACRO{\unit{GHz}}%
%BeginExpansion
\mathrm{GHz}%
%EndExpansion
$)\ to the antisymmetric case. By comparing these calculated resonances to the
directly measured resonances of B2 resonator, obtained using a broadband
$S_{11}$ measurement ($2.5152%
%TCIMACRO{\unit{GHz}}%
%BeginExpansion
\mathrm{GHz}%
%EndExpansion
,$ $4.425%
%TCIMACRO{\unit{GHz}}%
%BeginExpansion
\mathrm{GHz}%
%EndExpansion
,$ $6.3806%
%TCIMACRO{\unit{GHz}}%
%BeginExpansion
\mathrm{GHz}%
%EndExpansion
,$ $8.176%
%TCIMACRO{\unit{GHz}}%
%BeginExpansion
\mathrm{GHz}%
%EndExpansion
$)$,$ we find a good agreement between the two results. The missing resonances
do not get excited apparently because of the coupling location of the feedline
relative to the resonator.

\subsection{B3 resonator}

B3 resonator, in contrast, showed some larger discrepancy between the measured
value for the first mode $\sim1.6$ $%
%TCIMACRO{\unit{GHz}}%
%BeginExpansion
\mathrm{GHz}%
%EndExpansion
$ (seen in Fig. \ref{S11 two jumps}) and the theoretical value $f_{1}=2.4462$
$%
%TCIMACRO{\unit{GHz}}%
%BeginExpansion
\mathrm{GHz}%
%EndExpansion
$ calculated according to the approximated equation:
\begin{equation}
f_{n}=\frac{nc}{2l\sqrt{\varepsilon_{r}}}%
\end{equation}

where $n$ is the mode number, $c$ is the light velocity, $l$ is the
open-circuited line length ($\simeq20%
%TCIMACRO{\unit{mm}}%
%BeginExpansion
\mathrm{mm}%
%EndExpansion
),$ and $\varepsilon_{r}$ is the relative dielectric coefficient of the
sapphire ($\simeq9.4).$

%you can choose not to have a title for an appendix
%if you want by leaving the argument blank

%use section* for acknowledgement

\section*{Acknowledgment}

%optional entry into table of contents (if used)
%\addcontentsline{toc}{section}{Acknowledgment}
E.B. would especially like to thank Michael L. Roukes for supporting the early
stage of this research and for many helpful conversations and invaluable
suggestions. Very helpful conversations with Gad Eisenstein, Oded Gottlieb,
Gad Koren, Emil Polturak, and Bernard Yurke are also gratefully acknowledged.

%trigger a \newpage just before the given reference
%number - used to balance the columns on the last page
%adjust value as needed - may need to be readjusted if
%the document is modified later
%The "triggered" command can be changed if desired:

%references section
%NOTE: BibTeX documentation can be easily obtained at:
%http://www.ctan.org/tex-archive/biblio/bibtex/contrib/doc/

%can use a bibliography generated by BibTeX as a .bbl file
%standard IEEE bibliography style from:
%http://www.ctan.org/tex-archive/macros/latex/contrib/supported/IEEEtran/bibtex
%\bibliographystyle{IEEEtran.bst}
%argument is your BibTeX string definitions and bibliography database(s)
%<OR> manually copy in the resultant .bbl file
%set second argument of \begin to the number of references
%(used to reserve space for the reference number labels box)

%

%TCIMACRO{\TeXButton{TeX field}{\begin{biographynophoto}{Baleegh Abdo}
%} }%
%BeginExpansion
\begin{biographynophoto}{Baleegh Abdo}
%EndExpansion
(S'2002) was born in Haifa, Israel in 1979. He received the B.Sc. degree in
computer engineering, in 2002, and the M.Sc. degree in electrical engineering
in 2004, both from the Technion--Israel Institute of Technology, Haifa,
Israel. Currently he is pursuing the Ph.D. degree in electrical engineering at
the Technion. His graduate research interests are nonlinear effects in
superconducting resonators in the microwave regime, resonator coupling and
quantum computation.%
%TCIMACRO{\TeXButton{TeX field}{\end{biographynophoto}}}%
%BeginExpansion
\end{biographynophoto}%
%EndExpansion
%

%TCIMACRO{\TeXButton{TeX field}{\begin{biographynophoto}{Eran Segev-Arbel}} }%
%BeginExpansion
\begin{biographynophoto}{Eran Segev-Arbel}
%EndExpansion
was born in Haifa, Israel in 1975. He received the B.Sc. degree in electrical
engineering from the Technion--Israel Institute of Technology, Haifa, Israel,
in 2002. He is currentely working toward the MS.c. in electrical engineering
at the Technion. His research is focused on parametric gain in superconducting
microwave resonators.%
%TCIMACRO{\TeXButton{TeX field}{\end{biographynophoto}}}%
%BeginExpansion
\end{biographynophoto}%
%EndExpansion
%

%TCIMACRO{\TeXButton{TeX field}{\begin{biographynophoto}{Oleg Shtempluck}}}%
%BeginExpansion
\begin{biographynophoto}{Oleg Shtempluck}%
%EndExpansion
\textbf{\ }was born in Moldova in 1949. He received the M.Sc. degree in
electronic engineering from the physical department of Chernovtsy State
University, Soviet Union, in 1978. His research concerned semiconducters and
dielectrics. From 1983 to 1992, he was a team leader in the division of design
engineering in Electronmash factory, and from 1992 to 1999 he worked as stamp
and mould design engineer in Ikar company, both in Ukraine. Currently he is
working as a laboratory engineer in Microelectronics Research Center,
Technion- Israel Institute of Technology, Haifa, Israel.%
%TCIMACRO{\TeXButton{TeX field}{\end{biographynophoto}}}%
%BeginExpansion
\end{biographynophoto}%
%EndExpansion%
%TCIMACRO{\TeXButton{TeX field}{\newpage}}%
%BeginExpansion
\newpage
%EndExpansion
%

%TCIMACRO{\TeXButton{TeX field}{\begin{biographynophoto}{Eyal Buks}} }%
%BeginExpansion
\begin{biographynophoto}{Eyal Buks}
%EndExpansion
received the B.Sc. degree in mathematics and physics from the Tel-Aviv
University, Tel-Aviv, Israel, in 1991 and the M.Sc. and Ph.D. degrees in
physics from the Weizmann Institute of Science, Israel, in 1994 and 1998,
respectively. His graduate work concentrated on interference and dephasing in
mesoscopic systems. From 1998 to 2002, he worked at the California Institute
of Technology (Caltech), Pasadena, as a Postdoctoral Scholar studying
experimentally nanomachining devices. He is currently a Senior Lecturer at the
Technion---Israel Institute of Technology, Haifa. His current research is
focused on nanomachining and mesoscopic physics.%
%TCIMACRO{\TeXButton{TeX field}{\end{biographynophoto}}}%
%BeginExpansion
\end{biographynophoto}%
%EndExpansion

%You can push biographies down or up by placing
%a \vfill before or after them. The appropriate
%use of \vfill depends on what kind of text is
%on the last page and whether or not the columns
%are being equalized.

%\vfill

%Can be used to pull up biographies so that the bottom of the last one
%is flush with the other column.

%that's all folks

\end{document}